\documentclass[usenatbib]{mnras}
\usepackage{amsmath}
\usepackage{amsfonts}
\usepackage{amssymb}
\usepackage{graphicx}
\begin{document}

\pagerange{\pageref{firstpage}--\pageref{lastpage}} \pubyear{2017}

\title[Recurring starspot occultations on Qatar-2]{Recurring sets of recurring starspot occultations on exoplanet-host Qatar-2}

\author[T.~Mo\v{c}nik, J.~Southworth and C.~Hellier]{T.~Mo\v{c}nik,\thanks{E-mail: t.mocnik@keele.ac.uk} J.~Southworth and C.~Hellier\\
Astrophysics Group, Keele University, Staffordshire, ST5 5BG, UK}

\date{Accepted XXX. Received YYY; in original form ZZZ}

\maketitle

\label{firstpage}

\begin{abstract}
We announce the detection of recurring sets of recurring starspot occultation events in the short-cadence \textit{K2} lightcurve of Qatar-2, a K dwarf star transited every 1.34\thinspace d by a hot Jupiter. In total we detect 34 individual starspot occultation events, caused by five different starspots, occulted in up to five consecutive transits or after a full stellar rotation. The longest recurring set of recurring starspot occultations spans over three stellar rotations, setting a lower limit for the longest starspot lifetime of 58\thinspace d. Starspot analysis provided a robust stellar rotational period measurement of $18.0\pm0.2$\thinspace d and indicates that the system is aligned, having a sky-projected obliquity of $0\pm8^{\circ}$. A pronounced rotational modulation in the lightcurve has a period of $18.2\pm1.6$\thinspace d, in agreement with the rotational period derived from the starspot occultations. We tentatively detect an ellipsoidal modulation in the phase-curve, with a semi-amplitude of 18\thinspace ppm, but cannot exclude the possibility that this is the result of red noise or imperfect removal of the rotational modulation. We detect no transit-timing and transit-duration variations with upper limits of 15\thinspace s and 1\thinspace min, respectively. We also reject any additional transiting planets with transit depths above 280\thinspace ppm in the orbital period region 0.5--30\thinspace d.
\end{abstract}


\section{INTRODUCTION}

An obliquity is a misalignment angle between the stellar rotational and planet's orbital axis and is a key tracer of planet migration. Sky-projected obliquity can be measured spectroscopically by the Rossiter-McLaughlin effect \citep{Gaudi07}, and by Doppler tomography \citep{CollierCameron10} if a system is bright enough, or photometrically by tracing starspot occultation events \citep{Sanchis11}. According to the Transiting Extrasolar Planets Catalogue (TEPCat)\footnote{http://www.astro.keele.ac.uk/jkt/tepcat/tepcat.html} \citep{Southworth11}, sky-projected obliquity has been measured for a total of 100 systems, of which 85 host a transiting hot Jupiter (defined here as planets with orbital periods shorter than 10\thinspace d and masses between 0.3 and 13\thinspace $M_{\rm Jup}$). The discovery of the first hot Jupiter in 1995 triggered suggestions that such planets cannot form so close to their host stars but have instead formed beyond the ice line and migrated inwards to their current orbits \citep{Mayor95}. The planet formation and migration theories are far from well understood and expanding the sample of systems with known obliquities is of particular interest.

When a transiting planet occults a starspot it produces a bump of temporary brightening in the lightcurve \citep{Silva03}. The ability to detect starspot occultations is dependent on sufficient photometric precision and cadence of the observations, and on whether starspots are underneath a transit chord. Depending on the obliquity of the system the same starspot may be occulted in several consecutive transits or even after several stellar rotations.

Starspot occultations also provide a measurement of the stellar rotational period and add to our knowledge of starspots' lifetimes on stars other than Sun. Starspots' lifetimes are believed to be correlated with spot sizes and anticorrelated with supergranulae sizes \citep{Bradshaw14}. In the case of the Sun the sunspots' lifetimes range from hours to several months, with a median lifetime of less than a day \citep{Solanki03}, and a longest recorded lifetime of an individual sunspot of 137\thinspace d \citep{McIntosh81}. On other stars the longest reported starspot lifetime using the occultation technique is 100\thinspace d for the active main-sequence star Kepler-17 \citep{Desert11}, based on nearly continuous 17 months of \textit{Kepler} observations \citep{Borucki10}.

Since the failure of \textit{Kepler}'s second reaction wheel, its successor, \textit{K2}, suffers a reduced photometric precision due to the degraded pointing accuracy \citep{Howell14}. Despite the shortcomings, \textit{K2} continues to be able to detect starspots, as demonstrated by the detection of recurring starspot occultation events in the WASP-85 system \citep{Mocnik16}.

Qatar-2 (2MASS J13503740-0648145, EPIC 212756297) is a moderately bright $V=13.3$ K dwarf, hosting a hot Jupiter in a 1.34-d orbit which was discovered by \citet{Bryan12}. Multicolour photometry of five transits by \citet{Mancini14} revealed starspot occultations in all lightcurves with one potential occultation recurrence. The recurring occultation pair indicated a sky-projected obliquity of $4.3\pm4.5^{\circ}$. The rotational period was reported as $11.4\pm0.5$\thinspace d, which has since been corrected to $14.8\pm0.3$\thinspace d due to a calculation error \citep{Mancini16}. However, the analysis presented in this paper proves that the two starspot occultations analysed by \citet{Mancini14,Mancini16} were in fact caused by two different starspots and that even their corrected rotational period is incorrect.

In this paper we use the \textit{K2} short-cadence observations of Qatar-2, which reveals recurring sets of recurring spot crossings. Starspot analysis provided a robust determination of the stellar rotational period and a measurement of the sky-projected obliquity. We also measured the period of the pronounced rotational modulation, refined the system parameters, searched for transit-timing (TTVs) and transit-duration variations (TDVs), additional transiting planets and phase-curve variations. Lastly, we discuss the discrepancy between gyrochronological and isochronal age estimates.

After initial submission of this paper, \citet{Dai17} announced a similar paper analysing the same \textit{K2} data set. Their conclusions are largely in agreement with ours.

\section{\textit{K2} OBSERVATIONS}

Qatar-2 was observed during \textit{K2}'s Campaign 6 between 2015 July 14 and 2015 September 30 in the 1-min short-cadence observing mode. A continuous 79-d monitoring provided a total of 115\thinspace 890 images. We retrieved the target pixel file via the Minkulski Archive for Space Telescopes (MAST). The photometric extraction and spacecraft-drift artefact removal was carried out as done in \citet{Mocnik16}. In short, we used a 36-pixel fixed photometric aperture mask, temporarily removed the low-frequency stellar variability by normalizing the lightcurve with a low-order polynomial, and applied a self-flat-fielding (SFF) procedure using the Gaussian convolution of the measured normalized flux versus drift arclength. We then reintroduced low-frequency stellar variability by multiplying the corrected lightcurve with the same normalization polynomial as used earlier. The applied SFF procedure improved the median 1-min photometric precision from 1271 parts per million (ppm) to 854\thinspace ppm, which is within 10 per cent of the original \textit{Kepler} precision for a similarly bright star. After removal of all the quality-flagged data points, such as thruster firing events and cosmic rays, we retained 111\thinspace 807 data points. The lightcurve before and after the SFF procedure is shown in Fig.~1.

\begin{figure}
\includegraphics[width=8.3cm]{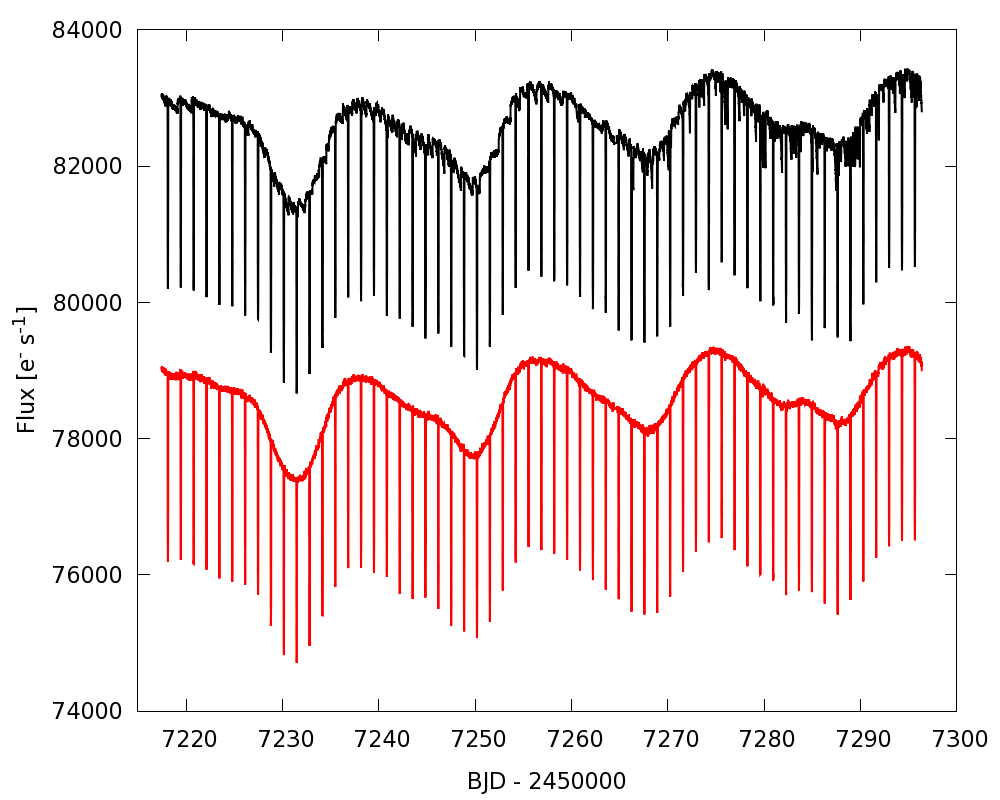}
\caption{Binned short-cadence lightcurve of Qatar-2 before (shown in black) and after the SFF correction (red). A 10-min binning was used for this plot to reduce the white noise and to display the drift artefacts more clearly. The lightcurve exhibits a total of 59 transits and a pronounced starspot rotational modulation. The corrected lightcurve is shown with an offset of \mbox{$-4000$ $\rm{e}^{-}\rm{s}^{-1}$} for clarity.}
\end{figure}

\section{SYSTEM PARAMETERS}

To determine the system parameters we performed a simultaneous Markov Chain Monte Carlo (MCMC) analysis of the \textit{K2} photometry coupled with the TRES radial-velocity measurements of Qatar-2 from \citet{Bryan12}. We re-applied the barycentric correction for TRES radial velocities to account for the issue that has been raised in their erratum where the original barycentric correction was calculated using the wrong declination. The corrected radial velocity measurements are available in Appendix~A. The MCMC code is presented in \citet{CollierCameron07} and further described in \citet{Pollacco08} and \citet{Anderson15}. Prior to the MCMC analysis we removed the starspot occultation events from the \textit{K2} lightcurve to prevent the potentially inaccurate parameter determination as discussed in \citet{Oshagh13}. We accounted for limb darkening by interpolating through tables of four-parameter limb-darkening coefficients, calculated for the \textit{Kepler} bandpass by \citet{Sing10}.

Because hot Jupiters are expected to circularise on a time-scale less than their age, we imposed a circular orbit to obtain the most likely parameters (e.g. \citealt{Anderson12}). To estimate the upper limit on eccentricity, we ran a separate MCMC analysis with eccentricity being fitted as a free parameter and checked the eccentricity distribution in the resulting MCMC chain.

To improve the precision of the orbital period, we produced a separate MCMC analysis that also included the ground-based photometry presented in \citet{Bryan12}, with limb-darkening coefficients from \citet{Claret00,Claret04}, as appropriate for different bandpasses. This extended the photometric baseline from 79\thinspace d to 5.5\thinspace yr and reduced the uncertainty to one fifth of its former value.

Table~1 lists the obtained system parameters and Fig.~2 shows the corresponding transit model.

We find a very good agreement with the system parameter values reported by \citet{Bryan12} with a significantly improved precision. \citet{Mancini14} reported stellar ($1.591 \pm 0.016$ $\rho_\odot$) and planetary densities ($1.183 \pm 0.026$ $\rho_{\rm Jup}$) significantly lower than \citet{Bryan12}. We find no evidence for the lower densities, as our values agree with those reported by \citet{Bryan12}.

\begin{figure}
\includegraphics[width=8.3cm]{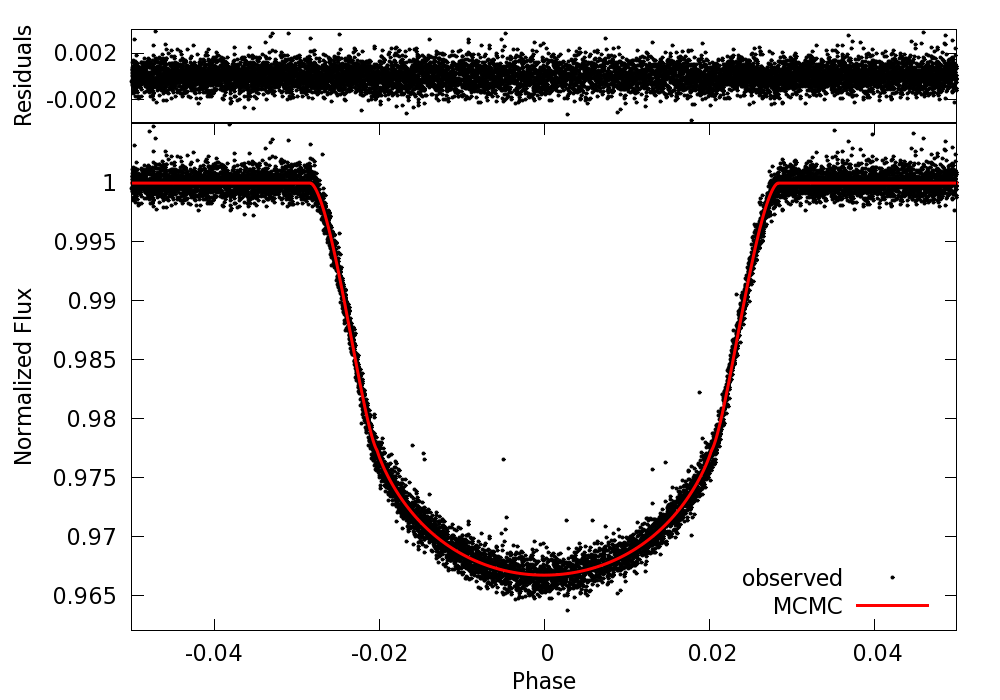}
\caption{Best-fitting MCMC transit model and its residuals from the phase-folded \textit{K2} lightcurve.}
\end{figure}

\begin{table*}
\centering
\begin{minipage}{11cm}
\caption{MCMC system parameters for Qatar-2 and Qatar-2b.}
\begin{tabular}{lccccc}
\hline
Parameter&Symbol&Value&Unit\\
\hline
Transit epoch&$t_{\rm 0}$&$2457250.2008155\pm0.0000084$&BJD\\
Orbital period&\textit{P}&$1.33711677\pm0.00000010$&d\\
Area ratio&$(R_{\rm p}/R_{\star})^{2}$&$0.026587\pm0.000062$&...\\
Transit width&$t_{14}$&$0.075409\pm0.000046$&d\\
Ingress and egress duration&$t_{\rm 12}$, $t_{\rm 34}$&$0.010696\pm0.000064$&d\\
Impact parameter&\textit{b}&$0.115\pm0.023$&...\\
Orbital inclination&\textit{i}&$88.99\pm0.20$&$^{\circ}$\\
Orbital eccentricity&\textit{e}&0 (adopted; $<$0.05 at $2\sigma$)&...\\
Orbital separation&\textit{a}&$0.02136\pm0.00024$&au\\
Stellar mass&$M_{\star}$&$0.727\pm0.024$&M$_\odot$\\
Stellar radius&$R_{\star}$&$0.7033\pm0.0080$&R$_\odot$\\
Stellar density&$\rho_{\star}$&$2.090\pm0.015$&$\rho_\odot$\\
Planet mass&$M_{\rm p}$&$2.466\pm0.062$&$M_{\rm Jup}$\\
Planet radius&$R_{\rm p}$&$1.115\pm0.013$&$R_{\rm Jup}$\\
Planet density&$\rho_{\rm p}$&$1.776\pm0.034$&$\rho_{\rm Jup}$\\
Planet equilibrium temperature$^{a}$&$T_{\rm p}$&$1285\pm16$&K\\
RV semi-amplitude&$K_1$&$0.5609\pm0.0063$&km\thinspace s$^{-1}$\\
Limb-darkening coefficients&$a_{\rm 1}$, $a_{\rm 2}$, $a_{\rm 3}$, $a_{\rm 4}$&0.703, $-0.737$, 1.486, $-0.642$&...\\
\hline
\end{tabular}
\begin{description}
\item[$^{a}$]Planet equilibrium temperature is based on assumptions of zero Bond albedo and complete day-to-night heat redistribution.
\end{description}
\end{minipage}
\end{table*}

\section{TTV AND TDV}

Measurements of mid-transit times and their deviations from a linear period ephemeris (transit-timing variations, or TTVs) can reveal additional and otherwise unobservable planets in planetary systems due to the inter-planet gravitational interactions \citep{Algol05}. Reported TTV amplitudes for perturbed transiting exoplanets are between a few tens of seconds and several hours, with typical periods of the order of a few hundred days \citep{Mazeh13}. Similarly, any additional perturbing planet would also cause transit-duration variations (TDVs) of the perturbed transiting planet, in phase with TTVs but with a significantly smaller amplitude \citep{Nesvorny13}.

We measured the transit-timings and transit-durations for Qatar-2b for each of the transits using the MCMC analysis as described in Section~3. Because the stellar activity can lead to inaccurate measurements of timings and durations \citep{Oshagh13}, we, as in Section~3, removed the detected starspot occultations from the lightcurve prior to the MCMC analysis. The resulting TTV and TDV measurements are shown in Fig.~3. The data point near BJD~2457269 has been excluded from the analysis due to the quality-flagged data gap (see transit number 39 in Fig.~5).

The TTV and TDV measurements yield $\chi^2$ values of 63.0 and 55.5, respectively, for 58 degrees of freedom, which means that they are statistically consistent with the assumption of white noise distributed around zero.  We estimate the semi-amplitude upper limit to be 15\thinspace s for TTVs and 1\thinspace min for TDVs. The absence of TTVs and TDVs indicates that a second, non-transiting, massive planet is unlikely, unless at a much longer period.

\begin{figure}
\includegraphics[width=8.3cm]{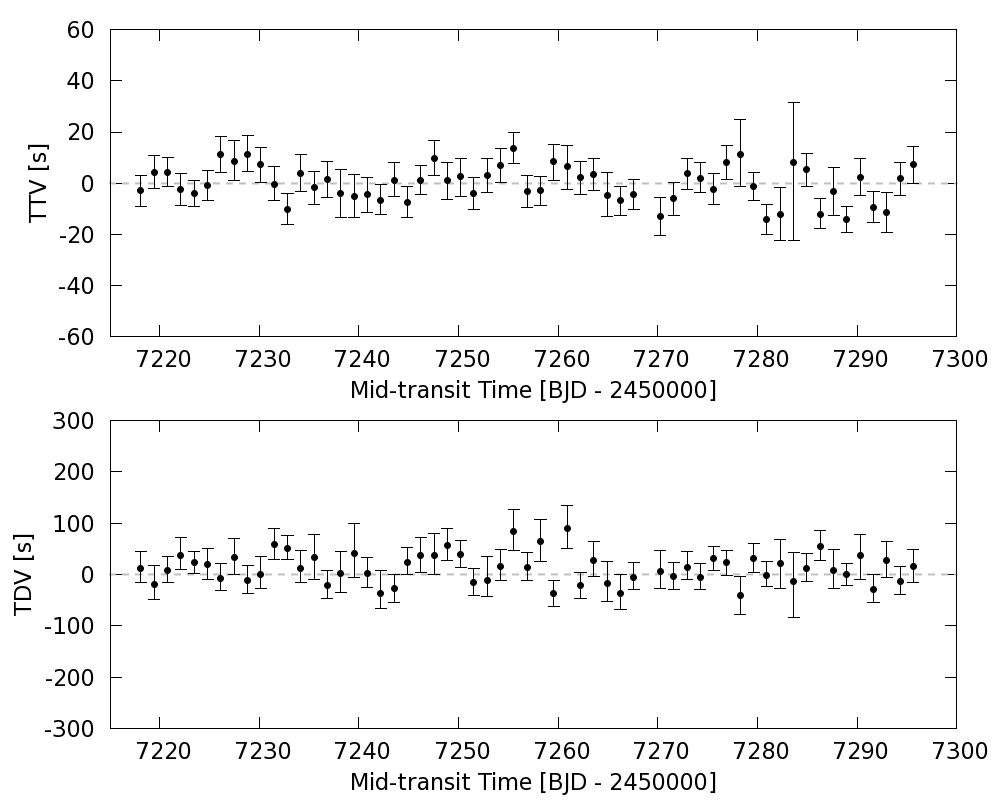}
\caption{The TTV (upper panel) and TDV (lower panel) measurements are statistically consistent with the assumption of strictly periodic transits and constant transit duration.}
\end{figure}

\section{ROTATIONAL MODULATION}

The \textit{K2} lightcurve of Qatar-2 exhibits a pronounced rotational modulation with an amplitude of about 2 per cent (see Fig.~1). The modulation is caused by the presence of starspots on the surface of the rotating host star. To measure the period we first removed the transits from the lightcurve, applied 10$\sigma$ clipping and then calculated a Lomb--Scargle periodogram. The rotational modulation period is manifested as the highest peak near 18.2\thinspace d with a Gaussian standard deviation of 1.6\thinspace d (see Fig.~4).

\begin{figure}
\includegraphics[width=8.3cm]{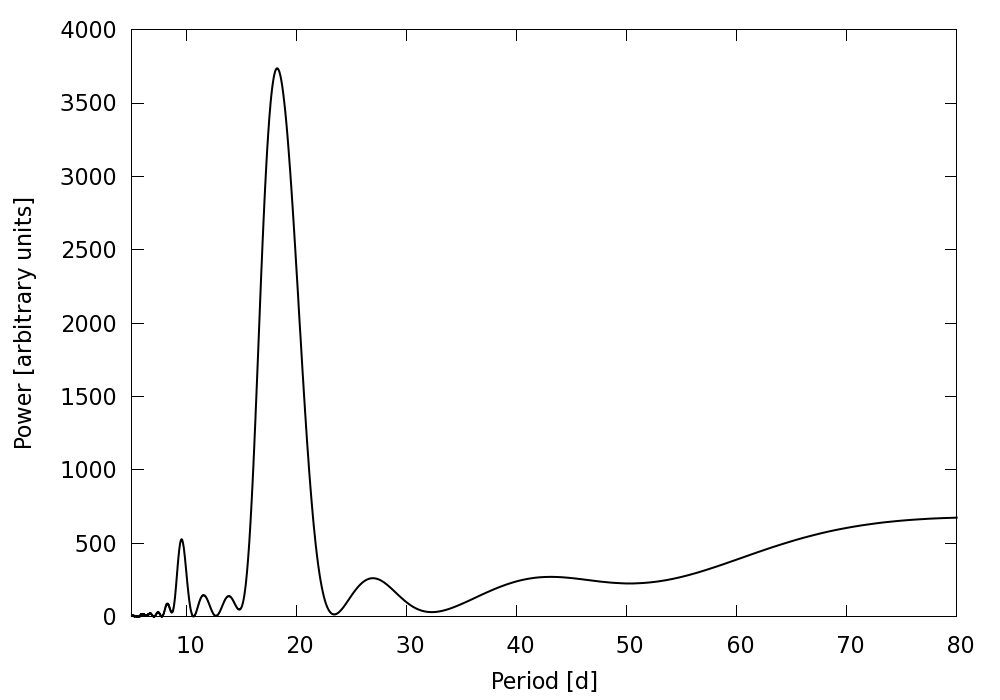}
\caption{Lomb--Scargle periodogram of the rotational modulation. The highest peak suggests the rotational period of $18.2\pm1.6$\thinspace d.}
\end{figure}

The resulting rotational modulation period of $18.2\pm1.6$\thinspace d is a direct measure of the rotational period. We measure a similar rotational period well within the uncertainty also when using three publicly available drift-corrected long-cadence lightcurves ({\scriptsize{K2SFF}} \citealt{Vanderburg14}, {\scriptsize{K2SC}} \citealt{Aigrain16} and {\scriptsize{EVEREST}} \citealt{Luger16}), which dismisses any doubts of whether our custom artefact removal technique might contaminate the rotational modulation.

The stellar rotational period given above and stellar radius given in Table~1 correspond to a stellar equatorial rotational velocity of $2.0\pm0.2$\thinspace km\thinspace s$^{-1}$, whereas the spectroscopically measured stellar projected rotational velocity is $2.0\pm1.0$\thinspace km\thinspace s$^{-1}$ \citep{Esposito17}. Although the calculated rotational velocity and the measured projected rotational velocity match, suggesting an edge-on stellar rotation, we can constrain the inclination only very weakly as being larger than $30^{\circ}$ owing to a large relative error bar for the projected rotational velocity. A stronger indication for a near-edge-on stellar rotation is revealed by the starspot position measurements in Section~6.3.

\section{STARSPOTS}

\subsection{Detection}

Fig.~5 shows the transit lightcurves after subtracting the best-fitting transit model from Section~3. The transits show many temporary brightenings that appear to be starspot occultation events.

We should first consider whether these could be residual artefacts from the \textit{K2} thruster firings. However, this is highly unlikely since: (1) we know the times of thruster firings, and these do not coincide with the starspot features; (2) the starspots recur in relation to the planetary orbital and stellar rotation cycles, which the spacecraft firings do not; (3) the drift artefacts have a step-like appearance, which the starspot features do not; and (4) spacecraft drifts were particularly consistent during the observing Campaign 6, which resulted in only minimal residual drift artefacts. Thus we can conclude that the features are indeed occultations of starspots.

In order to identify starspot occultation events in an unbiased way we shuffled the order of the transits in the residual lightcurve and then four of our colleagues examined them by eye and searched for any occultation-like features. Firstly, we considered lightcurve features as starspot occultation candidates only if they were marked as a possible occultation event by at least two colleagues. We then fitted each occultation candidate with a Gaussian function to determine the orbital phase at which it occurs, and its amplitude and width, and calculated the corresponding change in the Bayesian information criterion (BIC). Finally, we report here only those occultation candidates whose BIC change was greater than 6. These are marked with ellipses in Fig.~5 and listed in Table~2 along with stellar topocentric longitudes, which we calculated using the system parameters from Table~1 and the best-fitted occultation phases. Phase uncertainties in Table~2 are as quoted by the {\scriptsize{SciPy}} \citep{Jones01} Gaussian fits, and stellar longitude uncertainties were calculated from the phase and system parameters uncertainties.

In total we report the detection of 34 starspot occultation events, most of which appear at consistent phase shifts in up to five consecutive transits (transits 46--50). Furthermore, we found that sets of recurring occultations also appear at consistent phase shifts, in full agreement with the rotational modulation period (see Section~5) and the assumption that all recurring sets were produced by the same starspots. The occultation recurrence pattern and a very good fit to a single stellar rotational period (see Section~6.2) suggests that all the 34 individual starspot occultation events were caused by only five different starspots (each starspot is marked with a different colour in Fig.~5). In the longest recurring set of recurring occultations (transits 10--11, 23--25, 36--38 and 50--53), the starspot is seen during three full stellar rotations, spanning over 43 transits. This corresponds to a starspot lifetime of at least 58\thinspace d. Fig.~5 also reveals that the shapes of recurring occultation events are changeable, indicative of starspot evolution.

\begin{figure*}
\includegraphics[width=17.6cm]{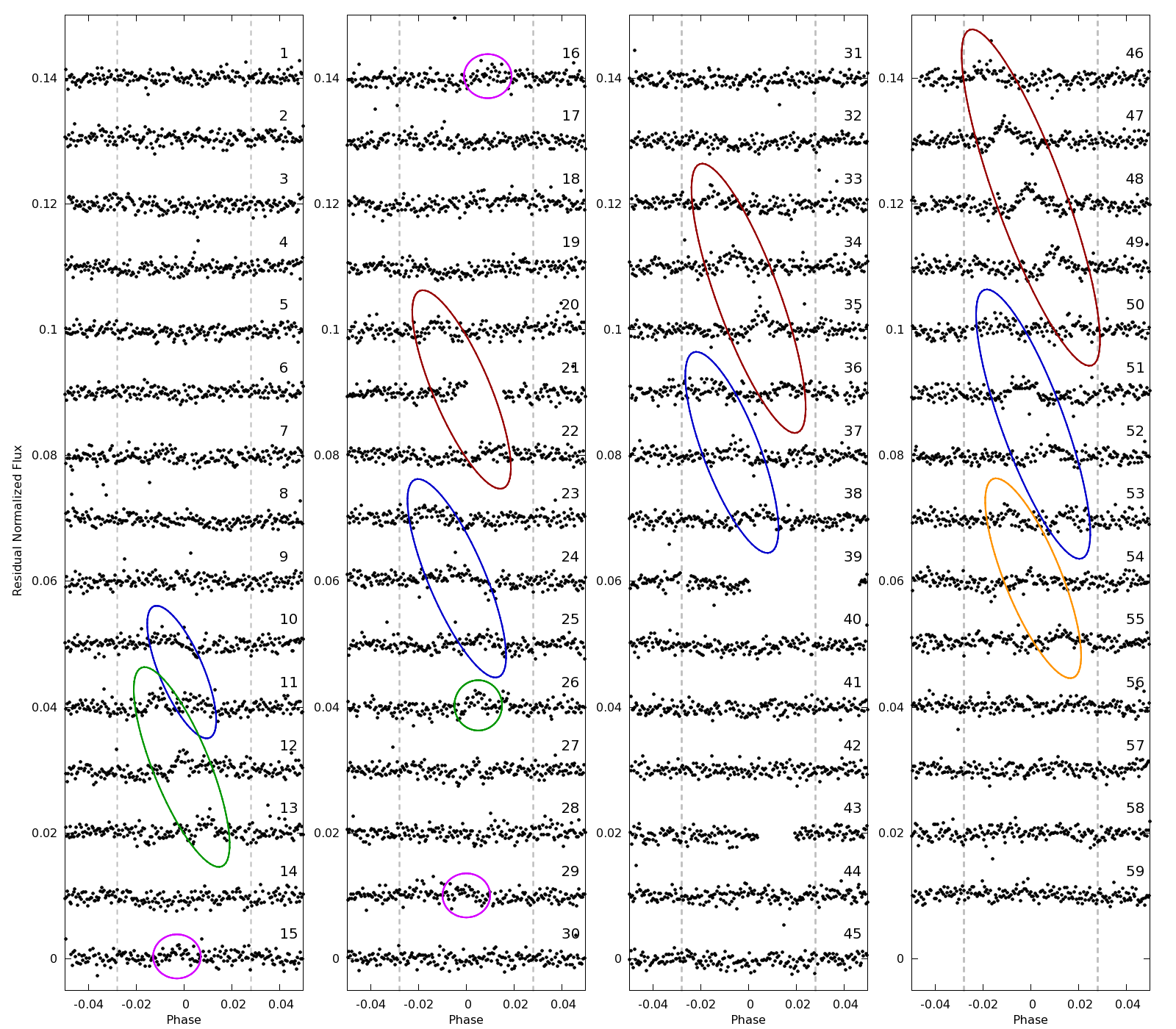}
\caption{Starspot occultations in the model-subtracted lightcurve of Qatar-2b. Vertical dashed lines specify the extent of the transit. Sets of recurring occultations are marked with ellipses. Ellipses of the same colour correspond to the same starspot.}
\end{figure*}

\begin{table}
\centering
\caption{Phase and stellar topocentric longitude positions of every detected starspot occultation event. The change of BIC provides an estimate of each occultation's detectability.}
\begin{tabular}{cr@{\thinspace $\pm$\thinspace}lr@{\thinspace $\pm$\thinspace}lc}
\hline
\multicolumn{1}{c}{Transit}&\multicolumn{2}{c}{Phase}&\multicolumn{2}{c}{Stellar}&\multicolumn{1}{c}{$\Delta$BIC}\\
\multicolumn{1}{c}{number}&\multicolumn{2}{c}{}&\multicolumn{2}{c}{longitude$^a$ ($^{\circ}$)}&\\
\hline
10&$-0.0087$&$0.0012$&$-21.1$&$3.0$&13.8\\
11&$-0.0112$&$0.0007$&$-27.5$&$1.9$&15.2\\
11&$0.0027$&$0.0010$&$6.4$&$2.4$&12.7\\
12&$-0.0006$&$0.0004$&$-1.4$&$1.0$&28.8\\
13&$0.0098$&$0.0005$&$23.9$&$1.3$&23.6\\
15&$-0.0025$&$0.0008$&$-5.9$&$1.9$&17.7\\
16&$0.0102$&$0.0014$&$24.9$&$3.7$&7.5\\
20&$-0.0128$&$0.0007$&$-31.9$&$1.9$&9.8\\
21&$-0.0012$&$0.0027$&$-2.8$&$6.4$&9.8\\
22&$0.0104$&$0.0012$&$25.4$&$3.2$&9.3\\
23&$-0.0155$&$0.0008$&$-39.8$&$2.4$&17.6\\
24&$-0.0049$&$0.0012$&$-11.7$&$2.9$&27.9\\
25&$0.0070$&$0.0009$&$16.8$&$2.2$&14.5\\
26&$0.0045$&$0.0005$&$10.7$&$1.2$&20.8\\
29&$-0.0010$&$0.0012$&$-2.4$&$2.8$&9.2\\
33&$-0.0151$&$0.0005$&$-38.6$&$1.5$&16.4\\
34&$-0.0062$&$0.0006$&$-14.8$&$1.5$&32.4\\
35&$0.0051$&$0.0003$&$12.2$&$0.7$&42.1\\
36&$-0.0141$&$0.0011$&$-35.6$&$3.1$&8.5\\
36&$0.0152$&$0.0008$&$38.9$&$2.5$&11.3\\
37&$-0.0060$&$0.0008$&$-14.3$&$2.0$&12.1\\
38&$0.0060$&$0.0007$&$14.3$&$1.7$&9.0\\
46&$-0.0185$&$0.0011$&$-49.8$&$3.9$&13.8\\
47&$-0.0106$&$0.0005$&$-26.0$&$1.3$&46.1\\
48&$-0.0008$&$0.0004$&$-1.9$&$1.0$&42.1\\
49&$0.0104$&$0.0010$&$25.4$&$2.7$&36.2\\
50&$-0.0147$&$0.0020$&$-37.4$&$5.7$&10.4\\
50&$0.0198$&$0.0011$&$54.8$&$4.8$&7.3\\
51&$-0.0028$&$0.0010$&$-6.6$&$2.4$&16.3\\
52&$0.0084$&$0.0007$&$20.3$&$1.8$&17.6\\
53&$-0.0096$&$0.0005$&$-23.4$&$1.3$&9.9\\
53&$0.0150$&$0.0012$&$38.3$&$3.7$&6.0\\
54&$0.0001$&$0.0010$&$0.2$&$2.4$&8.4\\
55&$0.0130$&$0.0010$&$32.5$&$2.9$&23.5\\
\hline
\end{tabular}
\begin{description}
\item[$^{a}$]Stellar topocentric longitude runs from $-90^{\circ}$ (first planetary contact), through $0^{\circ}$ (central meridian) to $90^{\circ}$ (last contact).
\end{description}
\end{table}

Although the rotational modulation shown in Fig.~1 is produced by a contribution from several detected starspots and likely also by some unocculted and therefore undetected starspots, we found a tentative correlation between the main modulation components and individual detected starspots. The starspot marked blue in Fig.~5 is seen near mid-transit close to the main rotational modulation's minima, suggesting that this was then the largest among Qatar-2's starspots. The occultation events marked red in Fig.~5 cluster near the plateaus or secondary minima (seen clearest near BJD 2457244 and 2457282 in Fig.~1). On the other hand, the rotational modulation's maxima occur when no or only small starspot occultation events are detected.

\subsection{Rotational period}

The stellar rotational period can be calculated from the changes in starspot positions on the rotating stellar surface and the timing of the occultation events. The position of the starspot is defined by stellar longitude and latitude and in general both parameters are needed for accurate determination of the starspot's position changes. However, in aligned systems such as Qatar-2 (see Section~6.3) and in the presence of recurring sets of recurring occultations, the most relevant starspot parameter is longitude.

Fig.~6 shows the measured phase and longitude positions of all the occultation events from Table~2. Using the best-fitting longitude vs. phase dependence for every recurring set, and the known orbital period of the planet, we were able to calculate the rotational period for every starspot. Their uncertainties were calculated from the slope uncertainties quoted by the {\scriptsize{SciPy}} fitting routine. The resulting rotational periods and their uncertainties are given in Table~3 and are based on the assumptions that starspots have circular shapes and that the starspots' positions change only due to stellar rotation, i.e. that starspot longitudinal migration is negligible. The latter assumption is justified by the solar observations which reveal that the longitudinal migration of sunspots is 4 orders of magnitude smaller than the solar rotation rate \citep{Gyenge14}. The weighted mean rotational period of $18.0\pm0.2$\thinspace d is in agreement with the rotational modulation period (see Section~5). We also performed a simultaneous fit of all recurring sets with a common rotational period and obtained the same result as a weighted mean. Because rotational modulations depend strongly on starspots coming and going, which causes modulational phase shifts, the period derived from starspot occultation events is more accurate and reliable than the period determined from the rotational modulation.

\begin{figure}
\includegraphics[width=8.3cm]{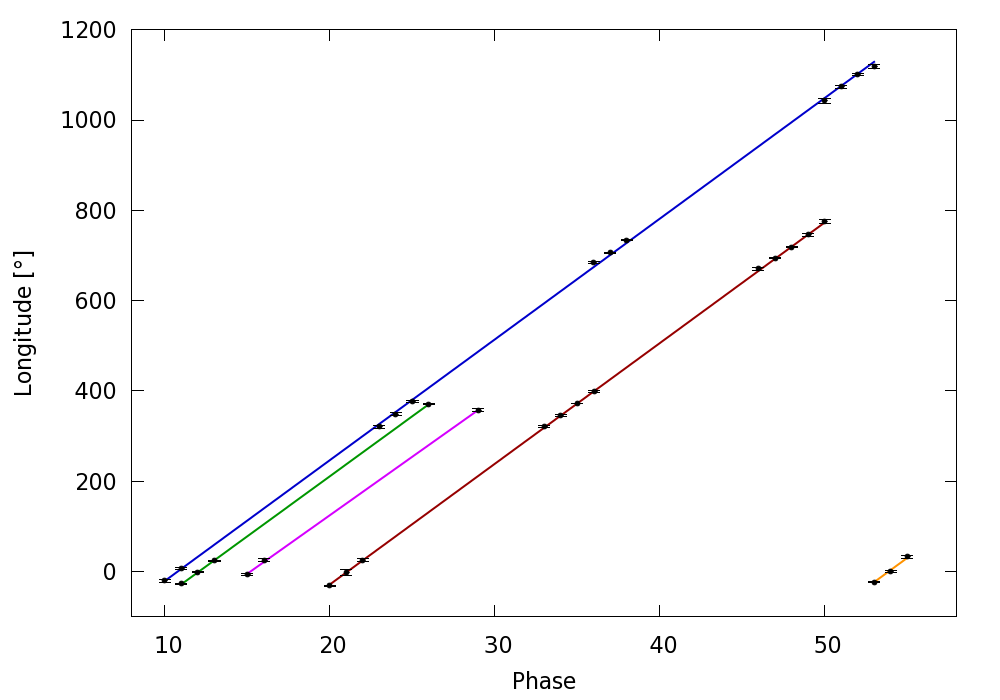}
\caption{Longitude and phase positions of all recurring occultations. We added $360^{\circ}$ to stellar longitudes from Table~2 for every stellar rotation. Each best-fitting straight line corresponds to a different starspot, which are marked with the same colours as in Fig.~5. The good fits to straight lines of very similar slopes suggests that occultation events of each recurring set were very likely caused by the same starspot. A few outliers and weak skewing trends, such as in the recurring set 50--53, may be attributed to non-circular and evolving starspot shapes, supported by the variable and non-Gaussian appearance of occultation events in Fig.~5.}
\end{figure}

\begin{table}
\centering
\caption{Stellar rotational period of every starspot.}
\begin{tabular}{cc}
\hline
Transit numbers&Rotational period (d)\\
\hline
10--11+23--25+36--38+50--53&$18.00\pm0.03$\\
11--13+26&$18.11\pm0.07$\\
15--16+29&$18.58\pm0.17$\\
20--22+33--36+46--50&$18.01\pm0.04$\\
53--55&$18.00\pm0.97$\\
\textit{weighted mean}&$\mathbf{18.03\boldsymbol{\pm}0.24}$\\
\hline
\end{tabular}
\end{table}

The recurring set of starspot occultations during transits 15--16 and 29 indicated a 0.5\thinspace d (2.3$\sigma$) longer rotational period than the weighted mean (see Table~3), possibly caused by a differential rotation. However, we were unable to investigate the effect of differential rotation due to the relatively small differences in rotational periods among different spots and because of poorly constrained starspot latitudes (see Section~6.3).

Our rotational period disagrees with the previously reported value of $14.8\pm0.3$\thinspace d by \citet{Mancini14,Mancini16}. Their rotational period was based on two starspot occultation events separated by 12\thinspace d, which were consistent with the projected rotational velocity of $2.8\pm0.5$\thinspace km\thinspace s$^{-1}$ by \citet{Bryan12}. With our robust rotational period of $18.0\pm0.2$\thinspace d it now seems reasonable to conclude that the two starspot occultation events analysed by \citet{Mancini14,Mancini16} (spots \#1 and \#2) were in fact caused by two different spots and that the projected rotational velocity by \citet{Bryan12} was overestimated by at least 1.5$\sigma$. A slower rotation than previously thought is further supported by the latest measurements of projected rotational velocity which place it at $2.0\pm1.0$\thinspace km\thinspace s$^{-1}$ \citep{Esposito17}.

\subsection{Obliquity}

The angle between the stellar rotational axis and the planetary orbital axis results in a gradual latitudinal positional drift of the recurring starspot occultation events. In misaligned systems the starspot occultations may occur only at preferential transit phases where the active latitudinal regions cross the planetary transit chord, such as in the HAT-P-11 system \citep{Sanchis11}. In the Qatar-2 system the starspot occultation events occur uniformly along the transit chord and in consecutive transits (see Fig.~5) which indicates that the system is aligned, as in the case of the Kepler-17 system \citep{Desert11}.

We fitted the starspot occultation events with the Planetary Retrospective Integrated Starspot Model ({\scriptsize{PRISM}}, \citealt{Tregloan13,Tregloan15}) and searched for any gradual starspot latitudinal drifts. We used the fixed input system parameters from Section~3 and searched for best-fitting starspot topocentric longitude, latitude, radius and contrast. We explored the entire starspot parameter space for each occultation event in each set of recurring occultations. Since {\scriptsize{PRISM}} was designed to model starspots with circular shapes, some of the occultation events in the Qatar-2 lightcurve were fitted poorly owing to their non-Gaussian shapes (e.g. set 50--53 indicated that the spot was elongated along the transit chord, possibly a strip of spots). We also found a high degree of degeneracy and non-physically rapid and non-monotonic starspot parameter variations among consecutive occultation events within several sets. Therefore, we decided to find the starspot radius and contrast that gave the best and most consistent fit within each set of consecutively recurring occultations and kept them fixed while refitting for starspot longitude and latitude positions. By using this approach we ignore starspot evolution, which may be correct only for time-scales significantly shorter than starspot lifetimes, and we assume that starspot contrast does not change as a function of the spot's distance from the stellar disc centre. The latter assumption is justified by the findings of \citet{Lanza03} who calculated that sunspot contrast changes only by a few per cent from the solar disc centre to the limb. Having the starspot radius and contrast fixed, the best-fitting latitudes could be determined from the widths and heights of each of the occultation events within the same set. This approach lifted the degeneracy and provided a physically more plausible solution.

In this paper we only present the obliquity results that were derived from the longest individual set of consecutively recurring occultations (transits 46--50), which yielded the most robust obliquity measurement with the smallest uncertainty. Firstly, this set exhibited pronounced Gaussian-like occultation events which enabled a reliable starspot parameter fitting. And secondly, by exhibiting five occultation events in consecutive transits it provided the largest longitudinal starspot coverage in the shortest possible time-span of 5\thinspace d, which is essential when assuming no starspot evolution.

The lightcurves and visualisations of the best-fitting starspot models for transits 46--50 are shown in Fig.~7. Starspot position parameters are listed in Table~4. The stellar longitude and latitude uncertainties in Table~4 are $1\sigma$ uncertainties as given by {\scriptsize{PRISM}}, which estimates uncertainties from the distribution of starspot parameters in the MCMC chain. We show the starspot longitudinal and latitudinal positions in Fig.~8 and find that they change linearly. If the stellar inclination was significant, the starspots' path would act as a quadratic function. We find no evidence for non-linear starspot paths which indicates that the stellar rotation must be close to edge-on. Using the system parameters from Section~3 and the fitted relation between the starspot longitudinal and latitudinal positions from Fig.~8, we derive the sky-projected obliquity of $-0.2\pm0.7^{\circ}$. For a degenerate but less-likely scenario in which a starspot appears below the transit chord, we obtain a similar sky-projected obliquity of $0.1\pm0.8^{\circ}$. Both calculations are based on the assumption that the starspot's size and contrast do not change throughout the set and that the starspot's latitudinal position only changes due to obliquity, i.e. neglects the contribution of latitudinal starspot migration. Therefore, the uncertainty is likely to be underestimated. In a very conservative approach of \citet{Desert11} we set the upper uncertainty limit to $8^{\circ}$, which is the largest possible sky-projected obliquity to still be able to see occultation events in five consecutive transits of Qatar-2b. The realistic uncertainty is expected to lie somewhere in between the two given values. However, because we were unable to quantify all the systematic errors and realistic ranges of starspot contrast and radius variability, we adopt $8^{\circ}$ as the final conservative uncertainty. A similar obliquity constraint was provided by \citet{Dai17} who claim that Qatar-2's sky-projected obliquity is smaller than $10^{\circ}$.

The small obliquity angle agrees with the empirical indication that systems with host stars cooler than 6250\thinspace K are generally aligned \citep{Winn10,Albrecht12}.

\begin{figure*}
\includegraphics[width=17.6cm]{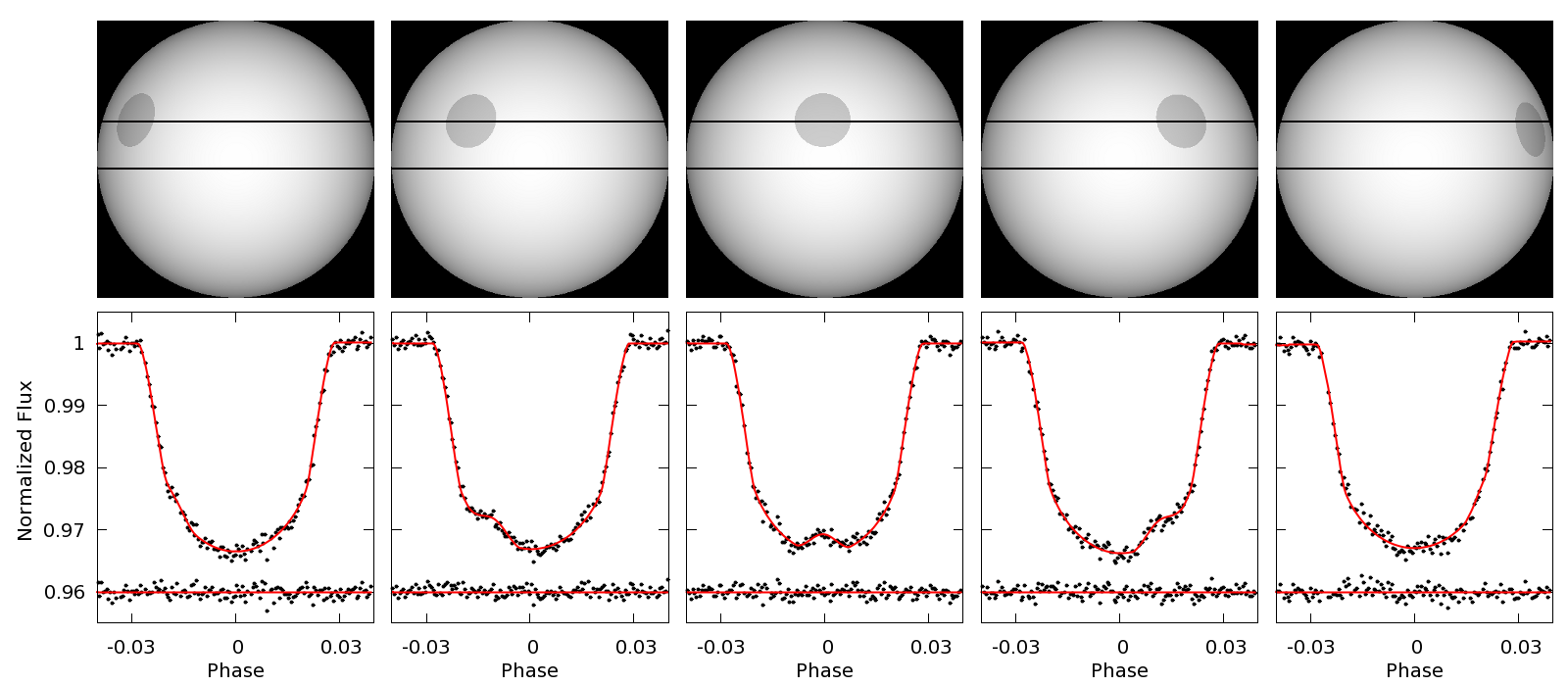}
\caption{Best-fitting starspot models for transits 46--50. Upper panels: starspot visualisations. Horizontal lines indicate the position of the transit chord. Lower panels: corresponding best-fitting lightcurve models. Shown below are the residual lightcurves after subtracting the starspot models. A notable residual in transit 50 at phase $-0.015$ is caused by another starspot (see Table~2 and Fig.~5).}
\end{figure*}

\begin{table}
\centering
\caption{Best-fitting starspot positions with {\scriptsize{PRISM}} for the longest individual set of recurring occultations, using the fixed starspot angular radius of $11.6\pm2.5^{\circ a}$ and contrast of $0.810\pm0.058^{b}$.}
\begin{tabular}{cr@{\thinspace $\pm$\thinspace}lc}
\hline
\multicolumn{1}{c}{Transit}&\multicolumn{2}{c}{Stellar}&\multicolumn{1}{c}{Stellar}\\
\multicolumn{1}{c}{number}&\multicolumn{2}{c}{longitude$^{c}$ ($^{\circ}$)}&\multicolumn{1}{c}{latitude$^{d}$ ($^{\circ}$)}\\
\hline
46&$-50.1$&$2.4$&$16.51\pm0.99$\\
47&$-26.8$&$1.1$&$16.15\pm0.66$\\
48&$-0.99$&$0.80$&$16.55\pm0.73$\\
49&$27.86$&$0.94$&$16.08\pm0.64$\\
50&$60.2$&$2.1$&$12.5\pm6.6$\\
\hline
\end{tabular}
\begin{description}
\item[$^{a}$]Angular radius runs from 0 to $90^{\circ}$, where $90^{\circ}$ covers half of the stellar surface. An angular radius of $11.6\pm2.5^{\circ}$ corresponds to an actual radius of $(9.9\pm2.1)\times10^4$\thinspace km.
\item[$^{b}$]Contrast runs from 0 to 1, where a value of 1 corresponds to the brightness of the surrounding photosphere.
\item[$^{c}$]Longitude runs from $-90^{\circ}$ (first planetary contact), through $0^{\circ}$ (central meridian) to $90^{\circ}$ (last contact).
\item[$^{d}$]Latitude runs from $-90^{\circ}$ (stellar south pole), through $0^{\circ}$ (equator) to $90^{\circ}$ (north pole).
\end{description}
\end{table}

\begin{figure}
\includegraphics[width=8.3cm]{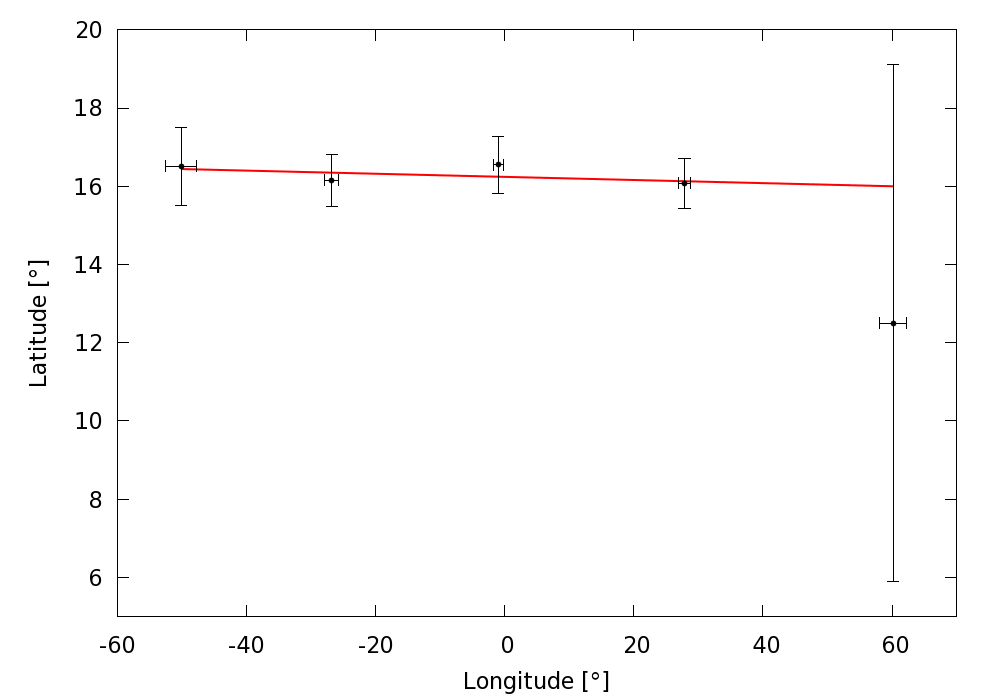}
\caption{Latitude and longitude positions for starspots in transits 46--50. The fit corresponds to the sky-projected obliquity angle of $-0.2\pm0.7^{\circ}$. The large uncertainty of the right-most data point (transit 50) is because of the starspot's proximity to the stellar limb and a low signal to noise of the occultation event.}
\end{figure}

\section{PHASE-CURVE MODULATION}

The three main phase-curve modulation components of planetary systems in the optical wavelengths covered by the \textit{K2} bandpass (420--900\thinspace nm) are: (1) reflection of starlight from an orbiting planet, (2) rotation of the star that is gravitationally distorted into an ellipsoidal shape, and (3) Doppler beaming of starlight by the orbital motion of the host star \citep{Esteves13}. To search for these phase-curve modulation components we first removed the rotational modulation and normalized the \textit{K2} lightcurve by dividing it with the median of second-order polynomial fits with window and step size of 3 and 0.3\thinspace d, respectively. Using this approach, we effectively removed the low-frequency rotational modulation while retaining any potential high-frequency phase-curve modulation, as confirmed by successful phase-curve signal injection and recovery tests. Next, we phase-folded the normalized lightcurve on the orbital period of Qatar-2b and applied binning to reduce the white noise (Fig.~9).

Using the system parameters from Table~1, the theoretically expected semi-amplitudes of the reflection, ellipsoidal and Doppler beaming phase-curve components are $595A_{\rm g}$\thinspace ppm, 12\thinspace ppm, and 8\thinspace ppm, respectively, where $A_{\rm g}$ is the planet's geometrical albedo \citep{Mazeh10}. We tentatively detect a sinusoidal signal peaking at orbital phases 0.25 and 0.75 with a minimum at phase 0.5, characteristic of ellipsoidal modulations. Its best-fitting semi-amplitude of $18\pm4$\thinspace ppm is close to the theoretically expected value. We were able to detect this modulation with a similar amplitude in the whole, first half and second half of the lightcurve, and also using three additional artefact removal procedures ({\scriptsize{K2SFF}}, {\scriptsize{K2SC}} and {\scriptsize{EVEREST}}). This strengthens the assumption that the detection of ellipsoidal modulation is real. However, the fit improves the reduced $\chi^2$ only slightly from 1.62 to 1.50. The lightcurve exhibits significant red noise such as SFF artefacts or residuals of the imperfect removal of the pronounced rotational modulation, and therefore this potential detection of ellipsoidal modulation should be regarded as tentative.

A reflectional modulation would peak at phase 0.5. A simultaneous MCMC fit of all three possible phase-curve modulation components gives the best-fitting reflectional modulation semi-amplitude of 2\thinspace ppm, though given the presence of red noise a conservative upper limit would be much higher, at 30\thinspace ppm. This reflection modulation upper limit of 30\thinspace ppm implies that the planet's optical geometric albedo has to be lower than 0.05. Low albedos at optical wavelengths are common for hot Jupiters \citep{Esteves13}, and are consistent with theoretical models which predict cloudless planetary atmospheres \citep{Burrows08}. The lowest optical geometric albedo has been measured for the TrES-2 system of less than 0.01 \citep{Kipping11}.

As of the time of writing, the phase-curve modulations have been detected for 26 planetary systems, 19 of which are from \textit{Kepler} observations according to NASA Exoplanet Archive\footnote{http://exoplanetarchive.ipac.caltech.edu/}. If our possible detection of ellipsoidal modulation is real, it would be the first phase-curve modulation detection of any of the planetary systems observed by \textit{K2}.

\begin{figure}
\includegraphics[width=8.3cm]{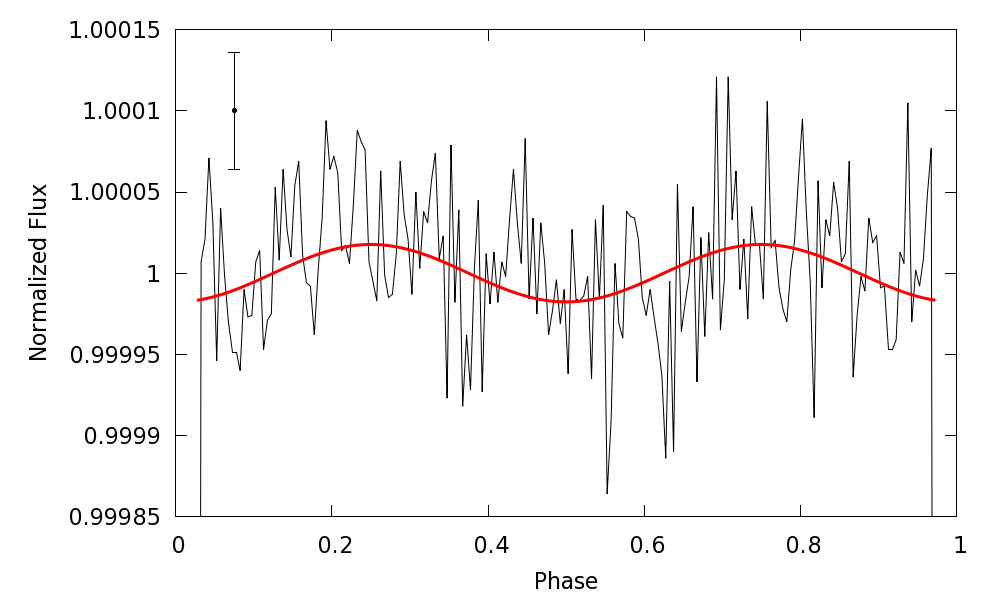}
\caption{Phase-curve of Qatar-2, binned to 200 bins. The red line denotes a possible detection of the ellipsoidal modulation with the best-fitting semi-amplitude of $18\pm4$\thinspace ppm. Rotational and Doppler beaming modulation components are not significantly detected. One representative error bar is shown in the top left corner.}
\end{figure}

\section{NO ADDITIONAL TRANSITING PLANETS}

We searched for any additional transiting planets in the normalized \textit{K2} lightcurve with low-frequency modulations removed. First, we removed Qatar-2b's transits from the lightcurve by replacing the measured normalized flux values around transits with unity. We then produced the periodogram to search for any remaining periodic transit-like features with {\scriptsize{PyKE}} tool {\scriptsize{KEPBLS}} which is based on the box-least-square fitting algorithm by \citet{Kovacs02}. The periodogram did not reveal any significant peaks between 0.5 and 30\thinspace d period with a transit depth upper limit of 280\thinspace ppm.

\section{AGE OF THE HOST STAR}

We estimate the stellar age first from the rotational period. Young stars rotate rapidly and gradually slow down as they lose angular momentum through magnetised stellar winds \citep{Barnes03}. We used the gyrochronological relation by \citet{Barnes07} to estimate that Qatar-2 is $0.59\pm0.10$\thinspace Gyr old.

Comparing stellar parameters to theoretical stellar models provides another age estimate. As in \citet{Mocnik16b} we compared stellar density and effective temperature of 4645\thinspace K to isochrones and evaluated the age of the star using the Bayesian mass and age estimator {\scriptsize{BAGEMASS}} \citep{Maxted15}. The stellar evolution models used in {\scriptsize{BAGEMASS}} were calculated using the {\scriptsize{GARSTEC}} code \citep{Weiss08}. The best-fitting stellar evolution track gives an age estimate of $9.4\pm3.2$\thinspace Gyr.

The huge discrepancy between gyrochronological and isochronal age estimates has also been seen in other exoplanet host stars, in particular K-type stars hosting a hot Jupiter \citep{Maxted15b,Mancini17}. It has been suggested that these discrepancies could be caused by the tidal interactions which could transfer angular momentum from an orbital motion of a planet to the rotation of a host star \citep{Maxted15b}. Thus, Qatar-2 may have been spun-up by the hot Jupiter and it is actually older than the gyrochronological age estimate. Another explanation for the age discrepancy is the radius anomaly of late type stars, where the observed stellar radii are larger than stellar model predictions \citep{Popper97}, which could result in isochronal ages being overestimated.

Obtaining a reliable age estimate for Qatar-2 would require additional age estimators such as the photospheric Li abundance, or an in-depth modelling to account for the two above-mentioned mechanisms that are probably contributing to the age discrepancy.

\section{CONCLUSIONS}

We used the \textit{K2} short-cadence observations of Qatar-2 from observing Campaign 6 to refine the system parameters, search for transit-timing variations, additional transiting planets, phase-curve modulations, and mainly to analyse the starspot activity.

The lightcurve of Qatar-2 exhibits pronounced rotational modulation with a period of $18.2\pm1.6$\thinspace d and an amplitude of about 2 per cent. The stellar activity is further manifested through 34 detected starspot occultation events among 59 transits. The majority of the occultations are seen repeatedly in up to five consecutive transits, owing to the low obliquity angle and a large ratio between stellar rotational and planet's orbital period. Moreover, sets of recurring occultations also reappear after several stellar rotations. The longest recurring set of recurring starspot occultations spans over three stellar rotations, which sets the longest starspot lifetime to at least 58\thinspace d. By fitting the recurring occultation events we derived the stellar rotational period of $18.0\pm0.2$\thinspace d, in agreement with the rotational modulation period, and zero sky-projected obliquity with a conservative error bar of $\pm8^{\circ}$.

We detect a possible ellipsoidal phase-curve modulation with a semi-amplitude of $18\pm4$\thinspace ppm, close to the theoretically expected value. Due to the presence of red noise this possible detection of ellipsoidal modulation should be regarded as tentative.

\section*{ACKNOWLEDGEMENTS}

We thank the anonymous referees for their perceptive comments which led to improving this paper. We gratefully acknowledge the financial support from the Science and Technology Facilities Council, under grants ST/J001384/1, ST/M001040/1 and ST/M50354X/1. This paper includes data collected by the \textit{K2} mission. Funding for the \textit{K2} mission is provided by the NASA Science Mission directorate. This work made use of {\scriptsize{PyKE}} \citep{Still12}, a software package for the reduction and analysis of \textit{Kepler} data. This open source software project is developed and distributed by the NASA Kepler Guest Observer Office. This research has made use of the NASA Exoplanet Archive, which is operated by the California Institute of Technology, under contract with the National Aeronautics and Space Administration under the Exoplanet Exploration Program.

\bibliographystyle{mnras}
\bibliography{bibliography}

\begin{thebibliography}{}
\makeatletter
\relax
\def\mn@urlcharsother{\let\do\@makeother \do\$\do\&\do\#\do\^\do\_\do\%\do\~}
\def\mn@doi{\begingroup\mn@urlcharsother \@ifnextchar [ {\mn@doi@}
  {\mn@doi@[]}}
\def\mn@doi@[#1]#2{\def\@tempa{#1}\ifx\@tempa\@empty \href
  {http://dx.doi.org/#2} {doi:#2}\else \href {http://dx.doi.org/#2} {#1}\fi
  \endgroup}
\def\mn@eprint#1#2{\mn@eprint@#1:#2::\@nil}
\def\mn@eprint@arXiv#1{\href {http://arxiv.org/abs/#1} {{\tt arXiv:#1}}}
\def\mn@eprint@dblp#1{\href {http://dblp.uni-trier.de/rec/bibtex/#1.xml}
  {dblp:#1}}
\def\mn@eprint@#1:#2:#3:#4\@nil{\def\@tempa {#1}\def\@tempb {#2}\def\@tempc
  {#3}\ifx \@tempc \@empty \let \@tempc \@tempb \let \@tempb \@tempa \fi \ifx
  \@tempb \@empty \def\@tempb {arXiv}\fi \@ifundefined
  {mn@eprint@\@tempb}{\@tempb:\@tempc}{\expandafter \expandafter \csname
  mn@eprint@\@tempb\endcsname \expandafter{\@tempc}}}

\bibitem[\protect\citeauthoryear{{Agol}, {Steffen}, {Sari}  \&
  {Clarkson}}{{Agol} et~al.}{2005}]{Algol05}
{Agol} E.,  {Steffen} J.,  {Sari} R.,   {Clarkson} W.,  2005, \mn@doi [\mnras]
  {10.1111/j.1365-2966.2005.08922.x}, \href
  {http://adsabs.harvard.edu/abs/2005MNRAS.359..567A} {359, 567}

\bibitem[\protect\citeauthoryear{{Aigrain}, {Parviainen}  \& {Pope}}{{Aigrain}
  et~al.}{2016}]{Aigrain16}
{Aigrain} S.,  {Parviainen} H.,   {Pope} B.~J.~S.,  2016, \mn@doi [\mnras]
  {10.1093/mnras/stw706}, \href
  {http://ukads.nottingham.ac.uk/abs/2016MNRAS.459.2408A} {459, 2408}

\bibitem[\protect\citeauthoryear{{Albrecht} et~al.,}{{Albrecht}
  et~al.}{2012}]{Albrecht12}
{Albrecht} S.,  et~al., 2012, \mn@doi [ApJ] {10.1088/0004-637X/757/1/18}, \href
  {http://adsabs.harvard.edu/abs/2012ApJ...757...18A} {757, 18}

\bibitem[\protect\citeauthoryear{{Anderson} et~al.,}{{Anderson}
  et~al.}{2012}]{Anderson12}
{Anderson} D.~R.,  et~al., 2012, \mn@doi [\mnras]
  {10.1111/j.1365-2966.2012.20635.x}, \href
  {http://adsabs.harvard.edu/abs/2012MNRAS.422.1988A} {422, 1988}

\bibitem[\protect\citeauthoryear{{Anderson} et~al.,}{{Anderson}
  et~al.}{2015}]{Anderson15}
{Anderson} D.~R.,  et~al., 2015, \mn@doi [A\&A] {10.1051/0004-6361/201423591},
  \href {http://adsabs.harvard.edu/abs/2015A%26A...575A..61A} {575, A61}

\bibitem[\protect\citeauthoryear{{Barnes}}{{Barnes}}{2003}]{Barnes03}
{Barnes} S.~A.,  2003, \mn@doi [\apj] {10.1086/367639}, \href
  {http://adsabs.harvard.edu/abs/2003ApJ...586..464B} {586, 464}

\bibitem[\protect\citeauthoryear{{Barnes}}{{Barnes}}{2007}]{Barnes07}
{Barnes} S.~A.,  2007, \mn@doi [\apj] {10.1086/519295}, \href
  {http://adsabs.harvard.edu/abs/2007ApJ...669.1167B} {669, 1167}

\bibitem[\protect\citeauthoryear{{Borucki} et~al.,}{{Borucki}
  et~al.}{2010}]{Borucki10}
{Borucki} W.~J.,  et~al., 2010, \mn@doi [Science] {10.1126/science.1185402},
  \href {http://adsabs.harvard.edu/abs/2010Sci...327..977B} {327, 977}

\bibitem[\protect\citeauthoryear{{Bradshaw} \& {Hartigan}}{{Bradshaw} \&
  {Hartigan}}{2014}]{Bradshaw14}
{Bradshaw} S.~J.,  {Hartigan} P.,  2014, \mn@doi [\apj]
  {10.1088/0004-637X/795/1/79}, \href
  {http://adsabs.harvard.edu/abs/2014ApJ...795...79B} {795, 79}

\bibitem[\protect\citeauthoryear{{Bryan} et~al.,}{{Bryan}
  et~al.}{2012}]{Bryan12}
{Bryan} M.~L.,  et~al., 2012, \mn@doi [\apj] {10.1088/0004-637X/750/1/84},
  \href {http://adsabs.harvard.edu/abs/2012ApJ...750...84B} {750, 84}

\bibitem[\protect\citeauthoryear{{Bryan} et~al.,}{{Bryan}
  et~al.}{2014}]{Bryan14}
{Bryan} M.~L.,  et~al., 2014, \mn@doi [\apj] {10.1088/0004-637X/782/2/121},
  \href {http://ukads.nottingham.ac.uk/abs/2014ApJ...782..121B} {782, 121}

\bibitem[\protect\citeauthoryear{{Burrows}, {Ibgui}  \& {Hubeny}}{{Burrows}
  et~al.}{2008}]{Burrows08}
{Burrows} A.,  {Ibgui} L.,   {Hubeny} I.,  2008, \mn@doi [\apj]
  {10.1086/589824}, \href {http://adsabs.harvard.edu/abs/2008ApJ...682.1277B}
  {682, 1277}

\bibitem[\protect\citeauthoryear{{Claret}}{{Claret}}{2000}]{Claret00}
{Claret} A.,  2000, A\&A, \href
  {http://adsabs.harvard.edu/abs/2000A%26A...363.1081C} {363, 1081}

\bibitem[\protect\citeauthoryear{{Claret}}{{Claret}}{2004}]{Claret04}
{Claret} A.,  2004, \mn@doi [A\&A] {10.1051/0004-6361:20041673}, \href
  {http://adsabs.harvard.edu/abs/2004A%26A...428.1001C} {428, 1001}

\bibitem[\protect\citeauthoryear{{Collier Cameron} et~al.,}{{Collier Cameron}
  et~al.}{2007}]{CollierCameron07}
{Collier Cameron} A.,  et~al., 2007, \mn@doi [MNRAS]
  {10.1111/j.1365-2966.2007.12195.x}, \href
  {http://adsabs.harvard.edu/abs/2007MNRAS.380.1230C} {380, 1230}

\bibitem[\protect\citeauthoryear{{Collier Cameron}, {Bruce}, {Miller}, {Triaud}
   \& {Queloz}}{{Collier Cameron} et~al.}{2010}]{CollierCameron10}
{Collier Cameron} A.,  {Bruce} V.~A.,  {Miller} G.~R.~M.,  {Triaud}
  A.~H.~M.~J.,   {Queloz} D.,  2010, \mn@doi [\mnras]
  {10.1111/j.1365-2966.2009.16131.x}, \href
  {http://adsabs.harvard.edu/abs/2010MNRAS.403..151C} {403, 151}

\bibitem[\protect\citeauthoryear{{Dai}, {Winn}, {Yu}  \& {Albrecht}}{{Dai}
  et~al.}{2017}]{Dai17}
{Dai} F.,  {Winn} J.~N.,  {Yu} L.,   {Albrecht} S.,  2017, \mn@doi [\aj]
  {10.3847/1538-3881/153/1/40}, \href
  {http://ukads.nottingham.ac.uk/abs/2017AJ....153...40D} {153, 40}

\bibitem[\protect\citeauthoryear{{D{\'e}sert} et~al.,}{{D{\'e}sert}
  et~al.}{2011}]{Desert11}
{D{\'e}sert} J.-M.,  et~al., 2011, \mn@doi [ApJS] {10.1088/0067-0049/197/1/14},
  \href {http://adsabs.harvard.edu/abs/2011ApJS..197...14D} {197, 14}

\bibitem[\protect\citeauthoryear{{Esposito} et~al.,}{{Esposito}
  et~al.}{2017}]{Esposito17}
{Esposito} M.,  et~al., 2017, \mn@doi [\aap] {10.1051/0004-6361/201629720},
  \href {http://ukads.nottingham.ac.uk/abs/2017A%26A...601A..53E} {601, A53}

\bibitem[\protect\citeauthoryear{{Esteves}, {De Mooij}  \&
  {Jayawardhana}}{{Esteves} et~al.}{2013}]{Esteves13}
{Esteves} L.~J.,  {De Mooij} E.~J.~W.,   {Jayawardhana} R.,  2013, \mn@doi
  [\apj] {10.1088/0004-637X/772/1/51}, \href
  {http://adsabs.harvard.edu/abs/2013ApJ...772...51E} {772, 51}

\bibitem[\protect\citeauthoryear{{Gaudi} \& {Winn}}{{Gaudi} \&
  {Winn}}{2007}]{Gaudi07}
{Gaudi} B.~S.,  {Winn} J.~N.,  2007, \mn@doi [\apj] {10.1086/509910}, \href
  {http://adsabs.harvard.edu/abs/2007ApJ...655..550G} {655, 550}

\bibitem[\protect\citeauthoryear{{Gyenge}, {Baranyi}  \&
  {Ludm{\'a}ny}}{{Gyenge} et~al.}{2014}]{Gyenge14}
{Gyenge} N.,  {Baranyi} T.,   {Ludm{\'a}ny} A.,  2014, \mn@doi [\solphys]
  {10.1007/s11207-013-0424-3}, \href
  {http://ukads.nottingham.ac.uk/abs/2014SoPh..289..579G} {289, 579}

\bibitem[\protect\citeauthoryear{{Howell} et~al.,}{{Howell}
  et~al.}{2014}]{Howell14}
{Howell} S.~B.,  et~al., 2014, \mn@doi [PASP] {10.1086/676406}, \href
  {http://adsabs.harvard.edu/abs/2014PASP..126..398H} {126, 398}

\bibitem[\protect\citeauthoryear{{Jones}, {Oliphant}  \& {Peterson}}{{Jones}
  et~al.}{2001}]{Jones01}
{Jones} E.,  {Oliphant} T.,   {Peterson} P.,  2001, {SciPy}: Open source
  scientific tools for {Python}, \url {http://www.scipy.org/}

\bibitem[\protect\citeauthoryear{{Kipping} \& {Bakos}}{{Kipping} \&
  {Bakos}}{2011}]{Kipping11}
{Kipping} D.,  {Bakos} G.,  2011, \mn@doi [\apj] {10.1088/0004-637X/733/1/36},
  \href {http://adsabs.harvard.edu/abs/2011ApJ...733...36K} {733, 36}

\bibitem[\protect\citeauthoryear{{Kov{\'a}cs}, {Zucker}  \&
  {Mazeh}}{{Kov{\'a}cs} et~al.}{2002}]{Kovacs02}
{Kov{\'a}cs} G.,  {Zucker} S.,   {Mazeh} T.,  2002, \mn@doi [\aap]
  {10.1051/0004-6361:20020802}, \href
  {http://adsabs.harvard.edu/abs/2002A%26A...391..369K} {391, 369}

\bibitem[\protect\citeauthoryear{{Lanza}, {Rodon{\`o}}, {Pagano}, {Barge}  \&
  {Llebaria}}{{Lanza} et~al.}{2003}]{Lanza03}
{Lanza} A.~F.,  {Rodon{\`o}} M.,  {Pagano} I.,  {Barge} P.,   {Llebaria} A.,
  2003, \mn@doi [\aap] {10.1051/0004-6361:20030401}, \href
  {http://ukads.nottingham.ac.uk/abs/2003A%26A...403.1135L} {403, 1135}

\bibitem[\protect\citeauthoryear{{Luger}, {Agol}, {Kruse}, {Barnes}, {Becker},
  {Foreman-Mackey}  \& {Deming}}{{Luger} et~al.}{2016}]{Luger16}
{Luger} R.,  {Agol} E.,  {Kruse} E.,  {Barnes} R.,  {Becker} A.,
  {Foreman-Mackey} D.,   {Deming} D.,  2016, \mn@doi [\aj]
  {10.3847/0004-6256/152/4/100}, \href
  {http://ukads.nottingham.ac.uk/abs/2016AJ....152..100L} {152, 100}

\bibitem[\protect\citeauthoryear{{Mancini} et~al.,}{{Mancini}
  et~al.}{2014}]{Mancini14}
{Mancini} L.,  et~al., 2014, \mn@doi [MNRAS] {10.1093/mnras/stu1286}, \href
  {http://adsabs.harvard.edu/abs/2014MNRAS.443.2391M} {443, 2391}

\bibitem[\protect\citeauthoryear{{Mancini} et~al.,}{{Mancini}
  et~al.}{2016}]{Mancini16}
{Mancini} L.,  et~al., 2016, \mn@doi [\mnras] {10.1093/mnras/stw1991}, \href
  {http://ukads.nottingham.ac.uk/abs/2016MNRAS.462.4266M} {462, 4266}

\bibitem[\protect\citeauthoryear{{Mancini} et~al.,}{{Mancini}
  et~al.}{2017}]{Mancini17}
{Mancini} L.,  et~al., 2017, \mn@doi [\mnras] {10.1093/mnras/stw1987}, \href
  {http://ukads.nottingham.ac.uk/abs/2017MNRAS.465..843M} {465, 843}

\bibitem[\protect\citeauthoryear{{Maxted}, {Serenelli}  \&
  {Southworth}}{{Maxted} et~al.}{2015a}]{Maxted15}
{Maxted} P.~F.~L.,  {Serenelli} A.~M.,   {Southworth} J.,  2015a, \mn@doi
  [\aap] {10.1051/0004-6361/201425331}, \href
  {http://adsabs.harvard.edu/abs/2015A%26A...575A..36M} {575, A36}

\bibitem[\protect\citeauthoryear{{Maxted}, {Serenelli}  \&
  {Southworth}}{{Maxted} et~al.}{2015b}]{Maxted15b}
{Maxted} P.~F.~L.,  {Serenelli} A.~M.,   {Southworth} J.,  2015b, \mn@doi
  [\aap] {10.1051/0004-6361/201525774}, \href
  {http://adsabs.harvard.edu/abs/2015A%26A...577A..90M} {577, A90}

\bibitem[\protect\citeauthoryear{{Mayor} \& {Queloz}}{{Mayor} \&
  {Queloz}}{1995}]{Mayor95}
{Mayor} M.,  {Queloz} D.,  1995, \mn@doi [\nat] {10.1038/378355a0}, \href
  {http://adsabs.harvard.edu/abs/1995Natur.378..355M} {378, 355}

\bibitem[\protect\citeauthoryear{{Mazeh} \& {Faigler}}{{Mazeh} \&
  {Faigler}}{2010}]{Mazeh10}
{Mazeh} T.,  {Faigler} S.,  2010, \mn@doi [\aap] {10.1051/0004-6361/201015550},
  \href {http://adsabs.harvard.edu/abs/2010A%26A...521L..59M} {521, L59}

\bibitem[\protect\citeauthoryear{{Mazeh} et~al.,}{{Mazeh}
  et~al.}{2013}]{Mazeh13}
{Mazeh} T.,  et~al., 2013, \mn@doi [ApJS] {10.1088/0067-0049/208/2/16}, \href
  {http://adsabs.harvard.edu/abs/2013ApJS..208...16M} {208, 16}

\bibitem[\protect\citeauthoryear{{McIntosh}}{{McIntosh}}{1981}]{McIntosh81}
{McIntosh} P.~S.,  1981, in {Cram} L.~E.,  {Thomas} J.~H.,  eds, The Physics of
  Sunspots. pp 7--54

\bibitem[\protect\citeauthoryear{{Mo{\v c}nik} et~al.,}{{Mo{\v c}nik}
  et~al.}{2016a}]{Mocnik16b}
{Mo{\v c}nik} T.,  et~al., 2016a, \mn@doi [\pasp]
  {10.1088/1538-3873/128/970/124403}, \href
  {http://ukads.nottingham.ac.uk/abs/2016PASP..128l4403M} {128, 124403}

\bibitem[\protect\citeauthoryear{{Mo{\v c}nik}, {Clark}, {Anderson}, {Hellier}
  \& {Brown}}{{Mo{\v c}nik} et~al.}{2016b}]{Mocnik16}
{Mo{\v c}nik} T.,  {Clark} B.~J.~M.,  {Anderson} D.~R.,  {Hellier} C.,
  {Brown} D.~J.~A.,  2016b, \mn@doi [\aj] {10.3847/0004-6256/151/6/150}, \href
  {http://adsabs.harvard.edu/abs/2016AJ....151..150M} {151, 150}

\bibitem[\protect\citeauthoryear{{Nesvorn{\'y}}, {Kipping}, {Terrell},
  {Hartman}, {Bakos}  \& {Buchhave}}{{Nesvorn{\'y}} et~al.}{2013}]{Nesvorny13}
{Nesvorn{\'y}} D.,  {Kipping} D.,  {Terrell} D.,  {Hartman} J.,  {Bakos}
  G.~{\'A}.,   {Buchhave} L.~A.,  2013, \mn@doi [ApJ]
  {10.1088/0004-637X/777/1/3}, \href
  {http://adsabs.harvard.edu/abs/2013ApJ...777....3N} {777, 3}

\bibitem[\protect\citeauthoryear{{Oshagh}, {Santos}, {Boisse}, {Bou{\'e}},
  {Montalto}, {Dumusque}  \& {Haghighipour}}{{Oshagh} et~al.}{2013}]{Oshagh13}
{Oshagh} M.,  {Santos} N.~C.,  {Boisse} I.,  {Bou{\'e}} G.,  {Montalto} M.,
  {Dumusque} X.,   {Haghighipour} N.,  2013, \mn@doi [\aap]
  {10.1051/0004-6361/201321309}, \href
  {http://adsabs.harvard.edu/abs/2013A%26A...556A..19O} {556, A19}

\bibitem[\protect\citeauthoryear{{Pollacco} et~al.,}{{Pollacco}
  et~al.}{2008}]{Pollacco08}
{Pollacco} D.,  et~al., 2008, \mn@doi [MNRAS]
  {10.1111/j.1365-2966.2008.12939.x}, \href
  {http://adsabs.harvard.edu/abs/2008MNRAS.385.1576P} {385, 1576}

\bibitem[\protect\citeauthoryear{{Popper}}{{Popper}}{1997}]{Popper97}
{Popper} D.~M.,  1997, \mn@doi [\aj] {10.1086/118552}, \href
  {http://adsabs.harvard.edu/abs/1997AJ....114.1195P} {114, 1195}

\bibitem[\protect\citeauthoryear{{Sanchis-Ojeda} \& {Winn}}{{Sanchis-Ojeda} \&
  {Winn}}{2011}]{Sanchis11}
{Sanchis-Ojeda} R.,  {Winn} J.~N.,  2011, \mn@doi [ApJ]
  {10.1088/0004-637X/743/1/61}, \href
  {http://adsabs.harvard.edu/abs/2011ApJ...743...61S} {743, 61}

\bibitem[\protect\citeauthoryear{{Silva}}{{Silva}}{2003}]{Silva03}
{Silva} A.~V.~R.,  2003, \mn@doi [\apjl] {10.1086/374324}, \href
  {http://adsabs.harvard.edu/abs/2003ApJ...585L.147S} {585, L147}

\bibitem[\protect\citeauthoryear{{Sing}}{{Sing}}{2010}]{Sing10}
{Sing} D.~K.,  2010, \mn@doi [A\&A] {10.1051/0004-6361/200913675}, \href
  {http://adsabs.harvard.edu/abs/2010A%26A...510A..21S} {510, A21}

\bibitem[\protect\citeauthoryear{{Solanki}}{{Solanki}}{2003}]{Solanki03}
{Solanki} S.~K.,  2003, \mn@doi [\aapr] {10.1007/s00159-003-0018-4}, \href
  {http://adsabs.harvard.edu/abs/2003A%26ARv..11..153S} {11, 153}

\bibitem[\protect\citeauthoryear{{Southworth}}{{Southworth}}{2011}]{Southworth11}
{Southworth} J.,  2011, \mn@doi [MNRAS] {10.1111/j.1365-2966.2011.19399.x},
  \href {http://adsabs.harvard.edu/abs/2011MNRAS.417.2166S} {417, 2166}

\bibitem[\protect\citeauthoryear{{Still} \& {Barclay}}{{Still} \&
  {Barclay}}{2012}]{Still12}
{Still} M.,  {Barclay} T.,  2012, {PyKE: Reduction and analysis of Kepler
  Simple Aperture Photometry data}, Astrophysics Source Code Library
  (\mn@eprint {ascl} {1208.004})

\bibitem[\protect\citeauthoryear{{Tregloan-Reed}, {Southworth}  \&
  {Tappert}}{{Tregloan-Reed} et~al.}{2013}]{Tregloan13}
{Tregloan-Reed} J.,  {Southworth} J.,   {Tappert} C.,  2013, \mn@doi [MNRAS]
  {10.1093/mnras/sts306}, \href
  {http://adsabs.harvard.edu/abs/2013MNRAS.428.3671T} {428, 3671}

\bibitem[\protect\citeauthoryear{{Tregloan-Reed} et~al.,}{{Tregloan-Reed}
  et~al.}{2015}]{Tregloan15}
{Tregloan-Reed} J.,  et~al., 2015, \mn@doi [MNRAS] {10.1093/mnras/stv730},
  \href {http://adsabs.harvard.edu/abs/2015MNRAS.450.1760T} {450, 1760}

\bibitem[\protect\citeauthoryear{{Vanderburg} \& {Johnson}}{{Vanderburg} \&
  {Johnson}}{2014}]{Vanderburg14}
{Vanderburg} A.,  {Johnson} J.~A.,  2014, \mn@doi [\pasp] {10.1086/678764},
  \href {http://ukads.nottingham.ac.uk/abs/2014PASP..126..948V} {126, 948}

\bibitem[\protect\citeauthoryear{{Weiss} \& {Schlattl}}{{Weiss} \&
  {Schlattl}}{2008}]{Weiss08}
{Weiss} A.,  {Schlattl} H.,  2008, \mn@doi [\apss] {10.1007/s10509-007-9606-5},
  \href {http://adsabs.harvard.edu/abs/2008Ap%26SS.316...99W} {316, 99}

\bibitem[\protect\citeauthoryear{{Winn} et~al.,}{{Winn} et~al.}{2010}]{Winn10}
{Winn} J.~N.,  et~al., 2010, \mn@doi [\apjl] {10.1088/2041-8205/723/2/L223},
  \href {http://adsabs.harvard.edu/abs/2010ApJ...723L.223W} {723, L223}

\makeatother
\end{thebibliography}

\appendix

\section{BARYCENTRIC-CORRECTED RADIAL VELOCITIES}
\citet{Bryan12} published 44 Qatar-2 radial velocity (RV) measurements which they obtained with the Tillinghast Reflector Echelle Spectrograph (TRES) between 2011 January 18 to 2011 June 21. As pointed out in their erratum \citep{Bryan14} the data reduction pipeline that they used to calculate the RVs had a bug, which parsed Qatar-2's negative declination with positive arcminutes and arcseconds for the barycentric correction. However, they did not provide the corrected RVs in their erratum. Table~A1 lists TRES RVs of Qatar-2 with correct barycentric correction.

\begin{table}
\centering
\caption{TRES barycentric-corrected RVs for Qatar-2. The RVs are given relative to the RV measured on BJD 2455646.85.}
\begin{tabular}{ccc}
\hline
BJD -- 2450000&RV (km~s\textsuperscript{-1})&$\Delta$RV (km~s\textsuperscript{-1})\\
\hline
5580.011622&0.001&0.034\\
5581.027117&0.158&0.038\\
5583.034587&$-0.796$&0.027\\
5583.961503&0.066&0.033\\
5585.003598&0.123&0.027\\
5587.983438&0.082&0.027\\
5602.958910&$-0.516$&0.033\\
5604.034894&0.097&0.030\\
5604.966229&$-0.051$&0.029\\
5605.966418&$-0.844$&0.024\\
5607.001348&$-0.638$&0.033\\
5607.987854&0.249&0.036\\
5608.955271&$-0.188$&0.028\\
5610.996605&$-0.533$&0.037\\
5615.938163&0.217&0.024\\
5616.973605&$-0.136$&0.029\\
5617.978176&$-0.860$&0.021\\
5643.881697&0.211&0.029\\
5644.849974&$-0.641$&0.026\\
5645.901592&$-0.803$&0.021\\
5646.846694&0.000&0.017\\
5647.890953&0.162&0.035\\
5650.857343&0.060&0.027\\
5652.895515&$-0.599$&0.033\\
5656.870962&$-0.706$&0.022\\
5659.877783&0.124&0.025\\
5662.839965&0.174&0.029\\
5663.812865&$-0.081$&0.021\\
5664.867650&$-0.763$&0.030\\
5665.796571&$-0.549$&0.021\\
5668.845188&$-0.838$&0.030\\
5671.762510&$-0.218$&0.022\\
5672.733966&$-0.858$&0.029\\
5673.784772&$-0.480$&0.019\\
5691.770974&$-0.340$&0.032\\
5702.699240&0.146&0.024\\
5703.727260&$-0.559$&0.032\\
5704.730193&$-0.816$&0.020\\
5705.694921&0.054&0.042\\
5706.738098&0.160&0.028\\
5722.744329&0.168&0.034\\
5726.718494&0.085&0.030\\
5728.755461&$-0.695$&0.032\\
5733.706680&0.189&0.028\\
\hline
\end{tabular}
\end{table}

\label{lastpage}

\end{document}